\documentclass[12pt]{article}
\usepackage{fullpage}
\usepackage{authblk}
\usepackage{natbib}

\title{Spectroscopic Binaries and Collapsed Stars: Part II}
\author[1,2]{Virginia Trimble}
\author[3]{Kip S. Thorne}
\affil[1]{\small{Department of Physics \& Astronomy, University of California, Irvine, CA 92697-4575}}
\affil[2]{Queen Jadwiga Observatory, Rzepiennik Biskupi, Poland}
\affil[3]{TAPIR (Theoretical Astrophysics Including Relativity), California Institute of Technology, Pasadena, CA 91125}

\begin{document}
\maketitle

\abstract{Fifty years ago, borrowing an idea from our Russian friends  Guseinov and  Zeldovich, we looked for very compact stars (neutron stars and black holes, in modern terminology) as optically-invisible components of single-line spectroscopic binaries. We didn't find any, but our method was very close to the processes that soon identified neutron stars in the X-ray sources Sco X-1, Cen X-3, and so forth, and the first persuasive black hole in Cyg X-1 (HD 226868). Here we look again at the events of 1962-72, revealing a bit more than we knew then, and attempt to bring the story up to date with an overview of some of the enormous richness of astronomical sources now generally thought to consist of a neutron star or black hole, or in a few cases one of each in a binary system.}

\section{Introduction}

John Michell (1724-1793) was almost certainly the first person to write about both binary stars and black holes, though not in those words and not in the same paper [\citet{michell1767} on double stars and clusters as physically real because they are too common on the sky to be chance superpositions; \citet{michell1784} on masses and distances of stars and also the possibility of escape speed exceeding the speed of light]. His sort of binary star was that where you see two dots of light in the sky, normally called visual binaries or double stars. We are not aware of any of these where a visible star orbits an invisible one of comparable mass.\\

Eclipsing binaries, one star passing periodically in front of the other from our point of view, include Algol, whose name suggests its oddity was known to the ancients. Only in 1782 did Edward Piggot and John Goodricke follow Algol carefully enough to suggest eclipses rather than dark spots on a single star as the cause of it variability. They backed off when an eclipsing model could not account for the light curve of Delta Cephei.\\

One hundred years then passed, bringing us to Hermann Carl Vogel at Potsdam and Edward C. Pickering at Harvard College Observatory, whose efforts to measure radial velocities of stars soon revealed points of light with two sets of spectral lines that moved around---spectroscopic binaries: Algol (confirming Goodricke \& Pigott's first thought) and Spica at Potsdam, and Mizar and Beta Aurigae (first spotted by Antonia Maury) at Harvard.	Since we shall not pass this way again, we pause to note that Vogel was called ``Geheimer Ober-Reg. Rath Professor Dr. Hermann Carl Vogel'' by S.~D. Townley \citep{townley1906} in awarding him the Bruce medal. Some of the things Pickering was called should probably not appear in this family-appropriate publication \citep{devorkin1999}.\\

Best of all for measuring stellar (including neutron stellar and black hole) masses and other properties are spectroscopic eclipsing binaries \citep{popper1967} (1913-1999). Double-lined SBs, or SB2s meaning that you see features from both stars, are best of all. Very few neutron star and black hole binaries rise to this standard, but you can do almost as well with two pulsars in mutual orbit (PSR J0737-3039) or a neutron-star rotation period for the second velocity amplitude (e.g.\ HZ Her = Her X-1).\\

Classical novae and their kin---recurrent novae, dwarf novae (SS Cygni and U Gem stars), nova-like variables, and so forth---are close binaries with a white dwarf as the accretor. This story was unfolding just before X-ray astronomy began, with binaricity established in 1949 \citep{sanford1949} for the recurrent nova T Cor Bor by R.~F. Sanford at Mt. Wilson (ret. 1950); for the dwarf novae SS Cyg and AE Aqu by \citet{joy1954, joy1956}; and for old novae by \citet{kraft1964}, but also \citet{kraft1962}.\\

Sanford followed the spectrum of T Cor Bor for several years after the (recurrent) nova event. It was an unholy mess of emission, absorption, and P Cygni-type lines. But as it faded, the absorption lines of a gM3 star became reliably visible/detectable. He measured radial velocities through what turned out to be 4 cycles with period 230.5 days and semi-amplitude $21 \, \mathrm{km/s}$ in a nice sine curve. There were also interstellar lines ($v = -29 \, \mathrm{km/s}$ about the central velocity of the M-star curve), a shell spectrum of hydrogen and such (scattered velocities), and emission lines including N II, O III, He II, Fe II, and forbidden [O III], [Ne III], [Ne V], and [Fe II]. These did not, however, provide a radial velocity curve anti-phased from the gM3 lines. Rather, they hovered around the systemic velocity, $-29 \, \mathrm{km/s}$, along with some (probably) circumstellar lines. Sanford therefore concluded that there was no evidence that T Cor Bor was a binary star! Well, OK, not one of those ideal SB2s.   But it \textit{was} a nice, clean SB1.\\

Roscoe Sanford has gone down in astro-history as one who was most congenial to his colleagues and who worked almost to his death, a decade after he retired (1883-1958). His BEA II \citep{hockey2014} entry \citep{trimble2014} mentions a number of achievements, especially contributions to establishing the existence of ``island universes'' (data used by Curtis in debates with Shapley), but not T Cor Bor---and we have no one to blame, since that entry was written by VT (p. 1916).\\

Bringing spectroscopic binaries and black holes (then called \textit{collapsed stars} or \textit{frozen stars}) closer together than had Michell was the idea of Oktay H. Guseinov, then working in Moscow with Zeldovich \citep{guseinov1966a, guseinov1966b, Zeldovich1966a}. Chapter 13 of \citet{Zeldovich1983} credits to Guseinov the idea that collapsed configurations would be easier to detect in binaries than alone, and \citet{Zeldovich1964a} had already proposed neutron stars as sources of accretion luminosity in the form of high-energy photons.\\

The 1966 \emph{Astronomicheskii Zhurnal} article \citep{guseinov1966a} put forward seven possible SBs with invisible massive companions. None of them is a strong X- or gamma ray source to this day, but several over the years turned out to be interesting in other ways.\\

The pair of papers \citet{Zeldovich1964b, Zeldovich1965e} [\citet{Zeldovich1965d, Zeldovich1966b} in English translation] talk about neutron stars, continued gravitational collapse, and collapsed or frozen stars (the latter comes out as ``cooled stars'' in translation, the more to confuse monolingual readers). There are some calculations of likely accretion rates and energy released, and citations of the \citet{oppenheimer1939a} and \citet{oppenheimer1939b} papers, of \citet{baade1934} on neutron star formation as a supernova energy source, and of \citet{landau1938}, on whatever you think that paper is about.\\

Over the next few years, at least five relevant things happened: (1) the number of known X-ray sources increased; (2) neutron stars as pulsars became part of the inventory of known objects (accepted very quickly, unlike black holes soon after); (3) \citet{batten1967} compiled a much more extensive and critical $8^{\mathrm{th}}$ catalogue of spectroscopic binaries with known orbits, (4) Kip Thorne visited Moscow and brought back the idea that compact objects might be detectable in SB1s; and (5) Trimble completed her Ph.D. dissertation at Caltech (on motions and structure of the filamentary envelope of the Crab Nebula) and had several stray weeks while the thesis was being typed, duplicated, and read by her committee before it could be defended (on 15 April 1968) and she could take off for England as a volunteer postdoc (paid for the summer, it turned out, as a librarian and receptionist).\\

KT suggested to VT that she should go through the Batten catalogue and identify all the SB1s with $M_2$ larger than $1.4 \, M_{\odot}$ and larger than the mass of the visible star $M_1$ as estimated from its spectral types (some of which went back to Annie J. Cannon!). Barbara~A. Zimmerman, the computing guru of Caltech's Kellogg Lab, provided essential help in writing and executing the program that took the catalogued properties (period and velocity amplitude) plus Kepler's third law and turned them into \begin{equation} \frac{M_2 \sin^3 i}{(M_1 + M_2)^2} = \mathrm{const.} \times P K_1^3, \end{equation} where $M_2$ is the mass of the invisible star, $M_1$ is the mass of the visible star that you have to guess from its spectral type, $P$ and $K_1$ are the period and the semi-amplitude in km/s or whatever, and $\sin i = 1$ yields a lower limit on $M_2$. This is customarily called the mass function, or $f(M)$.\\

We recovered six of the seven Zeldovich \& Guseinov stars and added six more with $M_2 > M_1$ and  $>1.4 M_\odot$, and another 38 with $M_2 < M_1$ but greater than $1.4 M_{\odot}$ (the limit for stable white dwarfs). Many years downstream, none of those systems is known to have a NS or BH component, and this was already obvious statistically at the time, because our search found about as many systems with $M_2 > M_1$ (\emph{etc.}) that showed eclipses (meaning that $M_2$ could not be a compact star) as that showed no eclipse.\\

Our paper was written during a two-day stopover VT made in Chicago, where KT was a visiting professor.   It is the only paper she ever handwrote (typing being so much easier). And there was a difficulty about who was to be first author. Not I, said VT, because it wasn't my idea. Not I, said KT, it wasn't mine either, it was the Russians'; and besides, I didn't do the necessary bits of arithmetic. Should it, he then asked, be Zeldovich \& Zimmerman? Well, perhaps not; and \citet{trimble1969} it then became, appearing first as a 1968 Orange Aid preprint (the arXiv of its day), and the page charges paid by Caltech, though VT was by then at Smith College.\\

Some of the tabulated systems weren't even binaries and did not appear in Batten's next, $9^{\mathrm{th}}$ catalogue, but many were interesting for other reasons, and our paper continued to be cited occasionally for many years.\\

Spectroscopic binaries remain today the kinds of things that Vogel and Pickering found; the collapsed stars for which we searched are today's astrophysical black holes, that is, celestial objects with radii on the order of a Schwarzschild radius, mass inconsistent with other known compact entities, and no peeking inside allowed. And general relativity has almost exactly doubled in age since we first considered the combination of spectroscopic binaries and collapsed stars.

\section{The Unconstrained Theorist}

Once upon a time, a theorist residing at infinity dropped some rotten tomatoes onto a white dwarf, a neutron star, and a black hole. Each gushy proton hitting the white dwarf surface with the escape speed would send out a 10 keV photon (an X-ray), or from the neutron star a 10 MeV gamma ray. The black hole just ate the tomato sauce and grew more massive. Then came accretion disks, because the theorist wasn't really at infinity, but on the surface of a companion star orbiting the compact object [\citet{shakura1973, pringle1972, prendergast1968, lynden-bell1969} though he was thinking primarily of supermassive black holes and quasars; \citet{novikov1973, kruszewski1966}].\\

The theoretical disk  provided ultraviolet from white dwarfs (as in the cataclysmic variables) and X-rays from the neutron stars. So far, so good. The white dwarf in the 165 year old nova V 841 Oph is still very bright in the UV---but also, it must be said, in soft X-rays \citep{cassatella1990}. And XRB without further qualification generally means ``with accreting neutron star.'' As for the black holes, especially those in galactic centers (perhaps even ours), it has been necessary to reintroduce ``down the tubes'' under the heading of Advection-Dominated Accretion.\\

It is perhaps worth noting that the G\&Z and T\&T searches took place when tomato­ behavior was not yet fully sorted out, and if we all focused mostly on X-ray sources (of which very few were yet known) it is because there were no generally-recognized gamma ray sources at all \citep{fazio1967}, while there were 21 X-ray sources that same year \citep{morrison1967}, some of which still exist with roughly the names and optical identifications suggested there.\\

\section{The Slightly Constrained Theorist}

Who-all should get primary credit for the idea that Sco X-1 was a binary system with a neutron star accreting from a companion has ceased to be angrily disputed, primarily because nearly all the claimants are no longer claiming. In T\&T we cited, for the general idea of accretion X- and gamma rays, \citet{salpeter1964} (more often mentioned in connection with black holes fueling QSRs), \citet{cameron1967}, and \citet{shklovsky1967}. Josef Samuilovich Shklovsky (1916-1985) is frequently credited as first past the post on this one, but the idea also appears in \citet{novikov1966}.\\

In the summer of 1966, there was an IAU Symposium in Noordwijk, NL \citep{iau1966}. One evening, outside the main program, a handful of men interested in the issue gathered to discuss Sco X-1 and such, with Geoffrey~R. Burbidge chairing. His version of the evening appears in \citet{burbidge1972}. He reports that Shklovsky (1916-1985) was there, but said not a word, while Vitaly Lazarevich Ginzburg was quite voluble.\\

Ginzberg (1916-2009), because he outlived the others and provided two late autobiographies \citep{ginzburg1990, ginzburg2001} got the last word. While claiming to have no interest in priority disputes, he recorded, first, that Zeldovich (1914-1981) had been annoyed that Shklovsky did not cite the Novikov \& Zeldovich paper from the previous year; and second, that he, Ginzburg, had the same binary neutron star idea independently and probably first, at or before Noordwijk. The three Russians also had ``issues'' over the recognition of astrophysical synchrotron radiation, and a few other high-energy astrophysics topics.  We enormously liked and deeply respected all three, and do not venture a vote.\\

Meanwhile, starting from the existence of considerable numbers of binary stars where the less massive component was the more evolved, theorists of a different mindset began exploring what happens when a pair of main sequence stars are close enough together that the attempts of the more massive to become a red giant end up transferring gas to the less massive. Breakthrough papers from three groups (in Munich, Prague, and Warsaw) date from 1967-1969, and by 1971 it was already possible to conceive of the process leading to neutron stars and black holes recieving material back from the initially less massive stars \citep{paczynski1971}. \Citet{vandenheuvel1973} bravely evolved a pair of intermediate-mass main sequence stars until one reached the black hole stage. (Yes, we are cited).

\section{Yes, Virginia (and Kip, and Oktay, and Yascha), There Is a Black Hole}

Indeed, it was lurking already in issues of \textit{Science}. Rocket flights in June 1964 and April 1965 \citep{bowyer1965, byram1966} recorded the faithful, steady Sco X-1 and also a source in Cygnus, whose flux relative to that of Sco dropped by a factor about four between the flights. Cyg X-1 existed, but it was not going to be a constant companion.\\

In December 1970, the Uhuru satellite launched (or \emph{was} launched, in our old-fashioned vocabulary). Cyg X-1 was there all right, but variable all over the time-scape \citep{oda1971, schreier1971}. A sudden flare-like spectral change on March 7 \citep{tananbaum1972b} sent radio observers to their telescopes to find a new or newly-brightened source, locatable to 0.5'' or so, inside the large X-ray error box \citep{wade1972, braes1971}. Armed with the radio position, optical astronomers quickly achieved an identification (to be called V something or other Cyg, you might suppose, but it was the previously­ catalogued star HDE 226868 \citep{bolton1972, webster1972}. The spectrum was that of a late O or early B supergiant (high temper­ature, low surface gravity) and variable radial velocity (half-amplitude about 75 km/s and period 5.6 days).\\

Give the primary a typical large mass for an OB supergiant, and the accretor must be at least $3.3 \, M_{\odot}$, which quickly grew to $6-10 \, M_{\odot}$ upon consideration of the likely orientation of the system. Clearly a black hole to any reasonable person, including, you might suppose, the four of us.\\

There was, nevertheless, a ``last wrong paper'' on the subject \citep{trimble1973}. It pointed out that post-AGB stars also go through a hot, low surface gravity phase, though at much lower luminosity than OB supergiants. They are also much less massive, and a visible star of that sort would have brought the accretor down into the neutron star mass range. Observers flocked to their telescopes, although flocks were smaller in those days \citep{bregman1973, margon1973} to prove us wrong by showing that HDE 226868 was 2-3 kpc away, not 200-300 pc, because its optical spectrum shows the scars of passing through a good deal of interstellar medium.\\

This was not a sad story, because our proposal was intended at least half in fun. The sad story dates from 1950, when \citet{popper1950} had included HDE 226868 in a program to measure radial velocities. He observed it twice, finding $V_r$'s different by only $4 \, \mathrm{km/s}$ (by contrast with the star's true  $\pm75$ km/s variations); and so  he reported it as non binary. Through extraordinarily bad luck, his two observations were made 263 days apart, a near-integral multiple of the orbital period.\\

Most of the clear spectroscopic binaries that Popper found he later followed up to get at least approximate orbits. Thus, had he recognized HDE 226868 as an SB1, it would probably have been in Batten's and others' catalogues to be included by G\&Z and T\&T as a candidate BH host system, at a position within the sizable error box of Cyg X-1.\\

In fact, Cyg X-1 eventually earned a variable star name, V1357 Cyg, and also has numerous X-ray and radio source catalogue identifiers, like 4U1956+35. VT heard the sad radial velocity story from Daniel Magnes Popper (1913-1999) on a long bus ride connected with some conference tour sometime between about 1975 and 1995.\\

Curiously, reviewers continued to write of ``black hole candidates'' for some years after we thought all the fuss was over \citep{eardley1975, bradt1983}.\\

By 1992, \cite{cowley1992} still wrote of ``candidates,'' but regarded Cyg X-1 as almost definitive and admitted LMC X-3, A0620-00 (= Nova Mon 1975, 1917), LMC X-1, and CAL 87 as good bets. In 1990 Hawking conceded to KT that Cyg X-1 is truly a black hole, in a bet since immortalized in the Hollywood movie \textit{The Theory of Everything}.  However, there are still living theorists who regard the evidence for black holes as unpersuasive,  as part of doubting, or rather firmly denying, the reality of gravitational waves; e.g.,  \citep{loinger2018}.

\section{Neutron Stars in X-Ray Binaries: Her X-1 and Sco X-1}

Sco X-1 was, of course, first.  It was discovered by Riccardo Giacconi's group at AS\&E, in a sounding rocket flight in June 1962 \citep{giacconi1962}.  Frank Paolini, the member of the discovery team least remembered in X-ray astronomy lore, built the detector that flew.  When their data showed a strong source in Scorpius, separated from the lunar positions they had intended primarily to scan, they got out a star atlas to see what else might be there. According to Herb Gursky,  they discovered there are a lot of stars in Scorpius, so they closed the star atlas, put it away, and concentrated on improving angular resolution [\citet{oda1965}---the modulation collimator]. \\

During their next flight, in October 1962,  again using Paolini's detector,  both the galactic center and Sco X-1 were below the horizon, so they saw no  strong X-rays. Sco X-1 was back in their third, June 1963 flight \citep{giacconi2010}, and the data showed it fixed in celestial coordinates.   Paolini immediately proclaimed Sco X-1 to be the first X-ray source outside the solar system. They had Air Force money for the flights because Bruno Rossi had urged AS\&E president Martin Annis to ask for a contract, and all three flights were on Aerobee 130 rockets.\\

An NRL flight \citep{bowyer1965} confirmed the Sco source. Lockheed also briefly entered the fray \citep{fisher1966}.  But it was Minoru Oda's modulation collimator that yielded a pair of error boxes small enough to suggest identification with an erratically variable, very blue star displaying emission lines rather like those from old novae. It is worth contemplating the much cited optical identification paper \citet{sandage1966} by 
A.~R.~Sandage, P.~Osmer (VT remembers being jealous of him in graduate school because he was a year behind her and got into that neat team), R. Giacconi P. Gorenstein, H. Gursky, T. Waters, H. Bradt, G. Garmire, B. V. Sreekantan, M. Oda, K. Osawa, and J. Jugaku.   The American authors  were all AAS members in 1973 (the earliest membership directory we have); Waters and Sreekantan were not, but the latter turns up at the Tata Institute in Bombay in the 1989 IAU Directory.\\

As further identifications turned up, \citet{burbidge1967} established Cyg X-2 as a binary (it was an NRL source), but Sco X-1 was so erratic in both visible and X-ray light that no orbit or other period could be established up to 1970 \citep{hiltner1970} from either radial velocity or photometric data. Uhuru came along, finding rotation periods for some of the X-ray components, which could be used as clocks to trace out orbital velocities, periodic in phase with the X-ray eclipses. Thus, Her X-1 and Cen X-3 became binary X-ray pulsars \citep{tananbaum1972a, schreier1972a}. It took Harvard College Observatory plates from 1890 to 1974 and a rather clever period-finding algorithm for \citet{gottlieb1975} to establish a photometric (orbital) period of 0.787 days or 18.9 hours for Sco X-1. Elaine W.\ Gottlieb  still has a Harvard e-dress, but not connected with astronomy. Finally, in 2002, \citet{steeghs2002} provided the first optical detection of the donor star in the form of emission lines from the irradiated side of the $0.42 \, M_{\odot}$ cool dwarf, though some correlation of optical and X-ray fluxes had confirmed the identification \citep{hudson1970}.\\

Her X-1 had the distinction of being the first X-ray source identified with a previously-known variable star. It was first picked out as significantly variable by \citet{hoffmeister1936} as one among ``604 Neue Ver\"{a}derliche.''  (Those 604, by the way, were among the 9,649 Hoffmeister variables out of 10,926 total from the inter-war Berlin-Babelsberg-Sonneberg Sky Patrol led by Paul Guthnick.) Hoffmeister called the star 142 1936. It became P 1483 as a provisional variable star designation ("kurzperiod") in Richard Prager's 1937 catalogue \citep{prager1937}. Though many international astronomical activities had been withdrawn from Germany with the 1919 establishment of the IAU, they retained variable stars, and we finally find HZ Her in the $39^{\mathrm{th}}$ namelist of variable stars \citep{guthnick1941}, where it is listed as RW Cnc type (with hearty thanks to Ron Webbink for these details!)\\

Prager was funded for a while by the inter-war IAU to continue systemization of variable stars, and he stuck to it a little too long, being imprisoned at Potsdam by the Nazis in 1938 for his Jewish ancestry---see \citet{robinson2014} for what happened next; not happy. Coming quickly down to the present, HZ Her appears in the \emph{Gaia} DR2 with periods of 1.2 s (rotation), 1.7 days (orbital), and 36 days (too complicated to explain here).\\

Her X-1 and Cyg X-1 are, of course, no longer unique as X-ray counterparts of previously-known stars. 4U0900-40 = HD 77581 and A0535+26 (an Ariel source) =  HDE 245770 and GS 2023+23 (from \citealp{makino1989}) was recognized by Brian Marsden as close to the position of V404 Cyg, a catalogued 1938 nova \citep{duerbeck1987}. The radio counterpart was announced by \citet{han1989}.

\section{A Sort of Timeline}

Most of the exciting things that happened between 1968-69 and 2018-19 had very little to do with spectroscopic binaries and collapsed stars (Biafra, Watergate, women clergy, AIDS, election of Nelson Mandela \ldots); others were clearly relevant but peripheral, like the 1990 debut of the World Wide Web. But our ``bookends'' here are a very early prediction that binary neutron stars would be X-ray sources \citep{hayakawa1964} and the discovery of a self-lensing binary star by \citet{kruse2014}. It is a white dwarf which magnifies its companion G star by 1 millimagnitude for 5 hours every 88 days\footnote{\texttt{https://www.dropbox.com/s/eb3pznxi990vn00/KOI3278\_pulse\_starspots2\_1024.mp4}}. That is to say, as our footnote in T\&T predicted, the effect is soberingly weak!\\

The years heading each item in the following timeline are those of discovery, and so often one ahead of the publications cited. 

1967. Discovery of pulsars \citep{hewish1968}. The Nobel Prize went to Hewish, not Bell, who had processed the original data and recognized something spectacularly new, an injustice that has wiped out thoughts of the injustice to the X-ray folks who had begun deliberately looking for neutron stars -- but an injustice perhaps rectified by Bell's 2018 Special Breakthrough Prize, with monetary value thrice larger than the Nobel.

1970. Launch of the AS\&E satellite Uhuru, followed by OSO-7 (1971), Copernicus \& SAS-2 (1972), ANS \& Ariel V (1974), SAS-3, OSO-8, \& COS-B (1975), HEAO-1 (1977), Einstein \& HEAO-2 (1978), GRS, Hachuko, \& HEA0-3 (1979), Tenma \& EXOSAT (1983), Ginga (1985), MIR/KVANT (1987), 1989 ROSAT \& GRANAT (1989), CGRO (1991), ASCA (1993), RXTE (1995), BeppoSAX (1996), Chandra (1999), and XMM-Newton (2000), these last two still operational. This list includes some missions that were primarily designated for gamma-ray astronomy, and is not complete. The satellite era permitted much deeper imaging (more sources), more accurate positions (more optical and radio IDs), and much longer timelines of observing (a whole zoo of kinds of variability). And someday we will have X-ray polarimetry!

1970. Correlation of optical and X-ray variations in Sco X-1 confirms the optical identification \citep{hudson1970}. Although Sco X-1 became quickly the prototypical XRB with a neutron-star accretor, spectral features attributable to the donor star were not detected until 2002 \citep{steeghs2002} via emission lines from the irradiated $0.42 \, M_{\odot}$ companion.

1972. Her X-1, the first X-ray source identified with a previously-known optically variable star \citep{tananbaum1972a}. Along with Cen X-3 (and unlike Sco X-1), it quickly revealed its binary nature via X-ray eclipses and Doppler effects yielding both orbital and rotation periods \citep{schreier1972a, schreier1972b}.

1974. PSR 1913+16, the first binary pulsar \citep{hulse1975}. Its relatively short (3 hour) orbital period and measurable eccentricity have made
it a playing field for tests of general relativity from	\citet{taylor1989} onward. It has passed all of those tests with flying colors, a not unmixed blessing, since many continue to hope for some hint toward an improved, quantum theory of gravity that might come from small failures of GR.

1975. Globular clusters reveal a large excess of binary X-ray sources (and in due course, binary millisecond pulsars, \citealt{clark1975}). Until then, it had been widely said that Population II stars included very few binaries, and this counterexample provided many theorists with an opportunity to explain the combination. VT dipped in here a bit later \citep{trimble1977} by not finding any main sequence eclipsing binaries in the globular cluster M55. The plates had been taken about 30 years earlier by Ivan~R. King to look for RR Lyrae stars. Attempted biomechanical blinking confirmed what King had said, ``first you think nothing varies; then you think everything varies; and then you actually have to go to work,'' and led to an attempt to use the APM (automated plate measuring machine) in Cambridge \citep{irwin1984}. Because very few astronomers love W UMa stars the way they should, there are perhaps still no good numbers for their incidence in globular clusters.

1975. A0620-00 \citep{elvis1975}. A spectacular X-ray transient spotted by Ariel V, this rose in real time during a European astronomy meeting in Leicester that summer, and was quickly revealed as a ``recurrent nova,'' having been optically birght in 1917 \citep{liller1975}.

1975.  Accretion luminosity can be augmented by nuclear reactions on the surface of the accretor, giving rise to Type I X-ray bursts in the case of neutron stars \citep{hansen1975} and classical novae in the case of white dwarfs.

1976. X-rays from SS433 \citep{seward1976}. It is also a radio source, catalogued as supernova remnant W50, unique in showing near-relativistic pre­cessing jets, for which there were soon many models. It has its own book \citep{clark1985}, and outflow from the jets may be as much responsible for W50 as the preceding supernova that presumably left the neutron star or black hole.

1981. The Fahlman-Gregory Object = CTB 109 \citep{fahlman1981}, an accretion-powered XRB in a supernova remnant, perhaps like SS433.

\item 1981. Sensitivities are such that a range of NS \& BH binary sources are seen in the LMC. There are no new types, but the statistics of type distribution is different from the Milky Way \citep{long1981}.	Types are now routinely described as having either NS or BH accretors (or pulsar non-accretors), and donors of high mass, low mass, or Be type (HMXRB, LMXRB, BeXRB). There seems to be a gap in the distribution of donor masses between about 2 and 8 $M_\odot$, and also a clean mass distinction between neutron stars of 1.2-1.44 $M_\odot$ and black holes of 6 $M_\odot$ or more.

1982. PSR 1937+11, the first millisecond (probably recycled, that is, spun up by accretion) pulsar \citep{backer1982}.

1985. Quasi-periodic Oscillations (QPOs) in the fluxes of (mostly) LMXRBs, an EXOSAT discovery \citep{vanderklis1985a, vanderklis1985b}. Within the next few years there came to be Z-sources, atoll sources, banana sources, and much else, as well as a garland of theories \citep{vanderklis1989}.

1986. X-rays from previously-catalogued stars no longer a rare phenomenon, and new optical IDs tend to add variable star names to their X-ray IDs: Vela XR-1 = HD77581, A0620-00 = V616 Mon, and so forth \citep{bradt1983}.

1987. Geminga = J0633+1746 \citep{bignami1988}, distinguished primarily by its very high ratio of gamma to X-ray (\emph{etc.}) luminosity.

1987. Statistics of millisecond pulsars and LMXRBs in globular clusters continue to puzzle theorists \citep{vandenheuvel1987}.

1988. The black widow pulsar 1957+20 \citep{fruchter1988}. The accretion-powered flux has eroded the donor star down to $0.02 \, M_{\odot}$ or thereabouts, and the system cannot keep this up much longer!

1989. GS 2023+23 \citep{makino1989}. It has a radio counterpart \citep{han1989}, and Brian Marsden found that it was catalogued as a nova, V404 Cyg, in 1938.

1992. PSR 1257+12 \citep{wolszczan1992}. The first pulsar with planets, so far the most persuasive case, and so the first extrasolar planets found from earth.

This is by no means the end of the discoveries, ideas, models, and so forth concerning neutron stars and black holes in stellar binary systems. Highlights in this territory, and many others, can be tracked for the next 16 years in \citet{trimble2007} and references cited therein. Beyond 2006, you are on your own.

\section{What Became of Us All?}

Yakov Borisovich Zeldovich (8 March 1914, Minsk - 2 December 1987, Moscow; and sporadically  Zel'dovich, though not in his own English-language signatures) had only recently moved from nuclear physics and hydrodynamics to astrophysics and cosmology in the early 1960s. He went on to a tremendous assortment of ideas, predictions, and accomplishments, for which see \citep{Scott2014} and references therein; \citet{ostriker1993}; and many other books, articles, and appreciations to be found with your favorite browser. The present authors were proud to count him a dear friend right up until his sudden death of a massive heart attack, which might not have been fatal in another medical system.\\

Oktay Huseyin Guseinov (elsewhere O.~Kh. Gusseinov, O.~H. Guseynov, and other spellings which do not cross-reference on ADS) is harder to trace. He was an Azerbaijani who worked with Zeldovich in Moscow---their 1965 ApJ paper was jointly submitted from the Institute of Physics of the Academy of Sciences of the USSR, Moscow;  but the Russian versions put him at Shemakha Observatory, Baku, Azerbaijan SSR \citep{guseinov1966a, guseinov1966b, Zeldovich1965a, Zeldovich1965c}, as indeed does his very first ADS paper, ``Neutronization of He$^4$'' \citep{Zeldovich1965b}. And there he remained, at least as an author, until 1991. The idea of searching for collapsed stars in SB1s came from him; Zeldovich wrote so in \cite{Zeldovich1983}. Other topics he worked on include white dwarfs, pulsars, supernova remnants, planetary nebulae, and ``experimental possibilities for observing cosmic neutrinosâ'' [a sole-author paper, \citet{guseinov1966c}] where he estimated that the neutrinos from a collapse to either neutron star or collapsed star---he first used the phrase ``black holeâ'' in 1973---would briefly drown out both solar neutrinos and cosmic ray secondaries. He cited papers by Bahcall, Pontecorvo, Reines, and Harrison, Thorne, Wakano \& Wheeler (that is, he was not working in complete isolation!) and thanked Zeldovich, Zatsepin, Kuz'min, Chudakov, and Domagatskii.\\

There are no Guseinov (\emph{etc.}) papers in ADS dated 1992, 93, or 94, and when he resurfaced in 1995, it was from T\"{U}BITAK in Antalya, Turkey (with his name printed in Turkish as Oktay H\"useyin'in \"Ozge\c cmi\c si). It was from there in the early 2000s that he turned up in email asking Trimble if she remembered him. Absolutely and affectionately, she e-bounced back immediately. The correspondence petered out, and we do not know whether he ever achieved the second or third cosmic velocities (Zeldovich's phrases) required to visit western Europe and the United States. Guseinov appears last as senior author of a 2006 edition of a Catalogue of X-ray Binaries \citep{guseinov2006}, and it was his colleagues on that project who posted the notice of his death on 23 March 2009.\\

Kip Thorne, previously a largish fish in a smallish pond of relativists, became world famous with his work on the Hollywood blockbuster \textit{Interstellar}. His fame expanded to other worlds (VT claims) when LIGO detected its first burst of gravitational waves, for which Thorne shared the 2017 Physics Nobel Prize---though, of course, it should  have gone to the entire LIGO team. He has remained at Caltech as professor emeritus, and today takes great joy in watching younger physicists create gravitational-wave astronomy while he himself mucks about in the interface between science and the arts. His book \emph{Modern Classical Physics}, co-authored with Roger Blandford (a 37 year labor of love), was recently published.\\

When our 1968/69 paper, T\&T, was written in Chicago, Trimble was on her way to spend a summer at Fred Hoyle's Institute of Theoretical Astronomy. After a year (1968-69) at Smith College as visiting assistant professor, she returned to Cambridge as a postdoc (1969-71), and then intended to settle into a tenure-track faculty position at the University of California, Irvine. But an encounter (February 1972) and marriage (March 1972) with Joseph Weber (1912-2000) led to their spending January to June each year at UCI and July to December at the University of Maryland, College Park. UMd fired Trimble in December 2003. Luckily her tenure resided at UCI, where she is now the oldest member of the Department of Physics \& Astronomy still on full active duty. She and Thorne were the youngest members of their departments when they started out! Trimble has received a handful of less prestigious (but very precious to her) forms of recognition,  most recently 
Asteroid 9271Trimble; 
Honorary Member, Astronomia Nova, Chestochowa Poland (2016);
the $15^{\mathrm{th}}$ William Tyler Olcott Distinguished Service Award of the American Association of Variable Star Astronomers; and designation as a Patron of the American Astronomical Society.\\

\vskip1pc

This paper was typed by VT, amended by KT, and keyboarded by Denise Schmidt, to whom we are enormously grateful.

\bibliographystyle{apj}
\bibliography{trimbleandthorne}

\end{document}